\documentclass[aps,prl,showpacs,notitlepage,amsmath,amssymb,floatfix,twocolumn,superscriptaddress]{revtex4-1}

\usepackage{amsmath} 
\usepackage{amssymb}
\usepackage{hyperref}
\usepackage[normalem]{ulem}
\usepackage{mathtools} 
\usepackage{mathptmx}

\DeclareMathAlphabet{\altmathcal}{OMS}{cmsy}{m}{n} 
\def\LN{\altmathcal{K}}                     % linking number
\def\CN{K}                                  % crossing number
\def\DH{\altmathcal{H}}                     % dimensionless helicity

\begin{document}

\title{Broken mirror symmetry of tracer's trajectories in turbulence}
\author{S.~Angriman}
\email[Corresponding author: ]{sangriman@df.uba.ar}
\affiliation{Universidad de Buenos Aires, Facultad de Ciencias Exactas y Naturales, Departamento de F\'\i sica, \& IFIBA, CONICET, Ciudad Universitaria, Buenos Aires 1428, Argentina}
\author{P.J.~Cobelli}
\affiliation{Universidad de Buenos Aires, Facultad de Ciencias Exactas y Naturales, Departamento de F\'\i sica, \& IFIBA, CONICET, Ciudad Universitaria, Buenos Aires 1428, Argentina}
\author{M.~Bourgoin}
\affiliation{Univ.~Lyon, ENS de Lyon, Univ.~Claude Bernard, CNRS, Laboratoire de Physique, 46 All\'ee d’Italie F-69342 Lyon, France}
\author{S.G.~Huisman}
\affiliation{Physics of Fluids Group, Max Planck UT Center for Complex Fluid Dynamics, Faculty of Science and Technology, MESA+ Institute and  J.M. Burgers Centre for Fluid Dynamics, University of Twente, P.O. Box 217, 7500 AE Enschede, The Netherlands}
\author{R.~Volk}
\affiliation{Univ.~Lyon, ENS de Lyon, Univ.~Claude Bernard, CNRS, Laboratoire de Physique, 46 All\'ee d’Italie F-69342 Lyon, France}
\author{P.D.~Mininni} 
\affiliation{Universidad de Buenos Aires, Facultad de Ciencias Exactas y Naturales, Departamento de F\'\i sica, \& IFIBA, CONICET, Ciudad Universitaria, Buenos Aires 1428, Argentina}

\begin{abstract} 
Topological properties of physical systems play a crucial role in our understanding of nature, yet their experimental determination remains elusive.
We show that the mean helicity, a dynamical invariant in ideal flows, quantitatively affects trajectories of fluid elements: the linking number of Lagrangian trajectories depends on the mean helicity. Thus, a global topological invariant and a topological number of fluid trajectories become related, and we provide an empirical expression linking them. The relation shows the existence of long-term memory in the trajectories: the links can be made of the trajectory up to a given time, with particles’ positions in the past. This property also allows experimental measurements of mean helicity.
\end{abstract}

\maketitle 
In recent years, broken symmetries and topology played an increasing role in physics. Examples are 
topological phase transitions \cite{Kosterlitz_2016}, topological charges in condensed matter \citep{Fradkin_2013}, applications in quantum field theory \cite{Kauffman_2001},  electromagnetism \cite{Kedia_2013}, DNA \cite{Vologodskii_1998} and chromosome organization \cite{Krepel2020}. In fluid dynamics, three-dimensional (3D) barotropic flows have an ideal invariant of topological nature. Helicity, the inner product between the Eulerian velocity $\mathbf{u}$ and the vorticity $\nabla\times\mathbf{u}$, integrated over the fluid volume $V$, 
$H = V^{-1} \int \mathbf{u}\cdot(\mathbf{\nabla}\times\mathbf{u})~dV $,
is proportional to the Gauss linking number of vorticity field lines \cite{Moffatt_1992, Moffatt_2014}, and measures their linkage and knottedness. Helicity is the only integral invariant of volume-preserving transformations 
\cite{Enciso2016}.
Moreover, 
its presence 
indicates the flow has no mirror symmetry (i.e., it is chiral).  Helicity is relevant in astrophysical \cite{Pouquet_1976, Yokoi_1993, Brandenburg_2005, Inagaki_2017} and geophysical flows \cite{Rasmussen_1998, Rorai_2013, Marino_2013}, in superfluids and Bose-Einstein condensates \cite{Bewley_2008, Rorai_2013b, Hall_2016, Tsatsos_2016, Clark_di_Leoni_2016, Kedia2018}, and in swirling \cite{Herbert2012} and rotating \cite{Mininni2009} flows. In active fluids, it can generate a helicity-driven inverse energy cascade (i.e., a self-similar transfer of energy to larger scales) \cite{Sahoo_2017}. In turbulence, the symmetry breaking introduced by non-zero helicity affects the statistical properties of the energy cascade, and leads to strongly depleted energy transfers between scales \cite{Kraichnan1988, Moffatt_2014}, or to a change in the energy transfer direction \cite{Sahoo_2017, Cameron2017}. 

Characterizing the topology of a vector field from a discrete set of measurements constitutes a cross-cutting challenge concerning several areas, such as surface reconstruction, deep learning, time series classification, and  chaotic attractor embeddings \cite{Wasserman_2018}. In fluid dynamics, helicity, although theoretically appealing, is hard to measure.  Experimental estimations employ pointwise measurements of velocity and vorticity (which are incomplete as helicity is a global quantity), or use linking numbers in flows simple enough that vorticity field lines can be identified \cite{Scheeler_2014, Scheeler_2017}. Helicoidal particles were also devised to estimate local flow chirality \cite{PhysRevFluids.1.054201}. Despite these attempts, measurements in the fully turbulent regime remain difficult, resulting on discussions on its conservation \cite{Kimura_2014, Laing_2015}.
 
Here we show that the broken mirror symmetry associated with helicity affects the connectivity of fluid elements trajectories, generating linkages between their long-time history. This is accomplished by combining simulations of homogeneous and isotropic turbulence (HIT) and of Taylor-Green (TG) flows at different Reynolds numbers, with laboratory experiments of mirror-symmetric HIT and of chiral von K\'arm\'an (VK) flows. The robustness of the results allows us to define a new volumetric measurement of helicity using the particle linking number, providing access to global quantification of helicity in experiments.

\begin{figure}[t] \centering
    \includegraphics[width=1\linewidth]{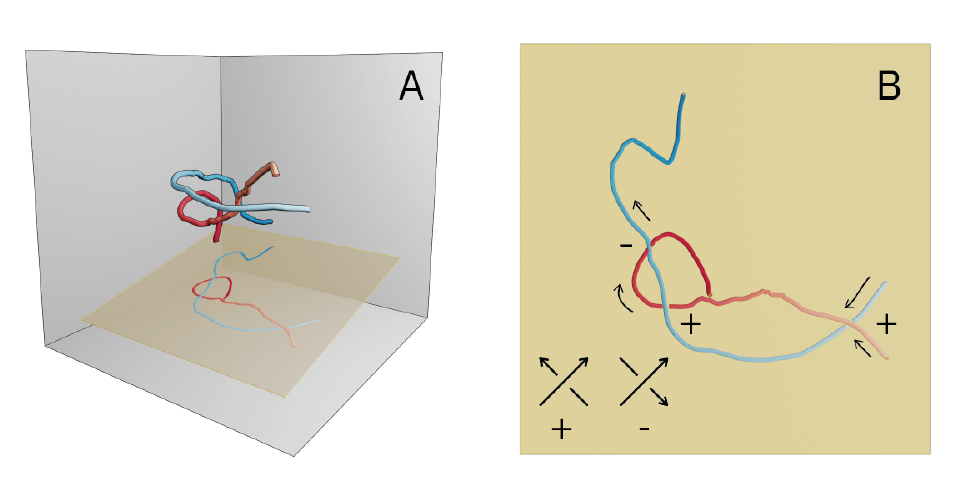} 
    \caption{ Definition of apparent crossing. A pair of experimental 3D trajectories (A) are projected onto a 2D plane (bottom of A, and B). Arrows and color gradients indicate time progression. A crossing is an intersection between the projections and is given by the particles' history; it does not necessarily occur with the particles being near at the same instant. The sign of each crossing is defined by the right-hand rule (B, bottom): $+1$ when an anti-clockwise rotation is needed to move from the tip of the arrow on top to the tip of the arrow below, and $-1$ in the clockwise case. Here, with two tracks projected in one plane, we see one negative and two positive crossings.}
    \label{fig:crossings}
\end{figure}

{\it Definition of the linking number of fluid elements’ trajectories.}
Does the number of links between tracers' trajectories constitute a proxy of the helicity of the underlying flow?
Particles' trajectories do not form, in general, closed loops. 
Even if some closed orbits exist under artificial (e.g., periodic) boundary conditions, laboratory measurements extend for a finite time and consist of short trajectories, spanning from a fraction to a few flow correlation times, so there is no notion of knottedness.
Still, we can define an average linking number between any set of 3D curves as the mean value of the signed apparent crossings in a $P$ number of two-dimensional (2D) projections. 
We thus consider a time interval $\Delta T$ over which we have measurements, and $N$ tracers trajectories that sample the flow during this interval. 
To compute the total number of signed crossings, we project the curves onto $P$ differently oriented planes, as if computing the 2D ``shadows'' of the trajectories. In each shadow, an apparent crossing  between two projected curves is defined as their intersection.
A crossing may also occur between two different time instants of the same trajectory; these self-crossings are treated identically
(we verified that removing self-crossings yields the same results). Note that crossings are not instantaneous crossings between particles, but between their history.
The sign of each crossing (i.e., counting it as $+1$ or $-1$) is given by the right-hand rule (see Fig.~\ref{fig:crossings}): we keep track of what trajectory is on top, and in what direction particles moved when going across the vertex. This orientation defines the handedness of the crossing \cite{Kauffman_2001}.

The normalized crossings $\CN_p$ for the $p$-th projection are defined as
$    \CN_p = M_p^{-1} \Sigma_{i=1}^{M_p}~\sigma_i$,
where $M_p$ is the total (unsigned) apparent crossings in the $p$-th projection, and $\sigma_i = \pm 1$ is the sign of the $i$-th crossing. Then, we define the mean linking number of the $N$ trajectories over the interval $\Delta T$ as the mean of $\CN_p$ over all $P$ projections: 
$\LN = P^{-1}\Sigma_{p=1}^{P} \CN_p$.

{\it Description of the data.} 
To study $\LN$ and $H$ we consider direct numerical simulations (DNSs) and tracers from particle tracking velocimetry (PTV) in laboratory experiments (see \cite{SI} for more details). We use two sets of DNSs \cite{Mininni_2011c, Rosenberg_2020} with resolutions of $256^3$, $512^3$, and $1024^3$ points to span different Reynolds numbers. The first consists of DNSs of HIT with correlated random forcing to give a flow with a tunable helicity \cite{Pouquet_1978} (we also consider a ``HIT 2'' simulation with very short forcing correlation time). Another set of DNSs uses TG forcing \cite{Brachet_1983, Ponty_2005} which mimics, in a periodic domain, multiple cells resembling VK flows, each non-mirror-symmetric and with non-null helicity (with alternating signs between the cells, resulting in null total helicity in a $(2\pi)^3$-periodic domain). 
In each simulation $10^6$ tracers were evolved along with the fluid. Experimental data on tracers trajectories obtained by PTV originates from two experiments: A VK experiment in Buenos Aires \cite{Angriman_2020} generates a helical flow, and the Lagrangian Exploration Module (LEM) in Lyon \cite{LEM, Bourgoin_2020} generates mirror-symmetric isotropic turbulence.

For each dataset, $\LN$ was computed using sets of trajectories that span a large-scale volume of the flow in non-overlapping time intervals $\Delta T$ ranging from a fraction to several $\tau_L$, with $\tau_L$ the Lagrangian correlation time (estimated from the tracers velocity auto-correlation function, or from structure functions in the LEM \cite{LEM,Sawford_2011}). 
We consider subsets of $N = 250$ particles in the DNSs and all available particles in the experiments, with $P = 26$ projections whose normal vectors are approximately uniformly distributed over a unit sphere, and given by
$\mathbf{\hat{n}}_p = (i \mathbf{\hat{x}} + j \mathbf{\hat{y}} + k \mathbf{\hat{z}})/(i^2 + j^2 + k^2)^{1/2}$ with $i,j,k \in \{-1, 0, 1\}$. 
This number of projections was empirically established as the minimum required to consistently recover the linking number of randomly oriented torus knots.

{\it 
The relation between the linking number of trajectories and helicity.}
We analyze $\LN$ as a function of the normalized, dimensionless
helicity $\DH$
\begin{equation}
    \DH = L U^{-2} \left< H \right>_{\Delta T},
    \label{eq:normalized_helicity}
\end{equation} 
where $\langle \cdot \rangle_{\Delta T}$ indicates time averaging 
over $\Delta T$, $U = (\langle  v_x^2 + v_y^2 + v_z^2\rangle )^{1/2}$ (with $v_i$ the components of the tracers' velocity and $\langle \cdot \rangle$ the average over $\Delta T$ and all trajectories) is a measure of the tracers' velocity over $\Delta T$ (to consider possible effects  of velocity variations over $\Delta T$ on the helicity), and $L = u \tau_L$ is a flow integral scale based on the characteristic one-component r.m.s.~value of the tracers' velocity $u$ estimated over a long time interval. These choices allow for estimation of all quantities solely from Lagrangian measurements.

\begin{figure} 
    \centering
    \includegraphics[width=1\linewidth]{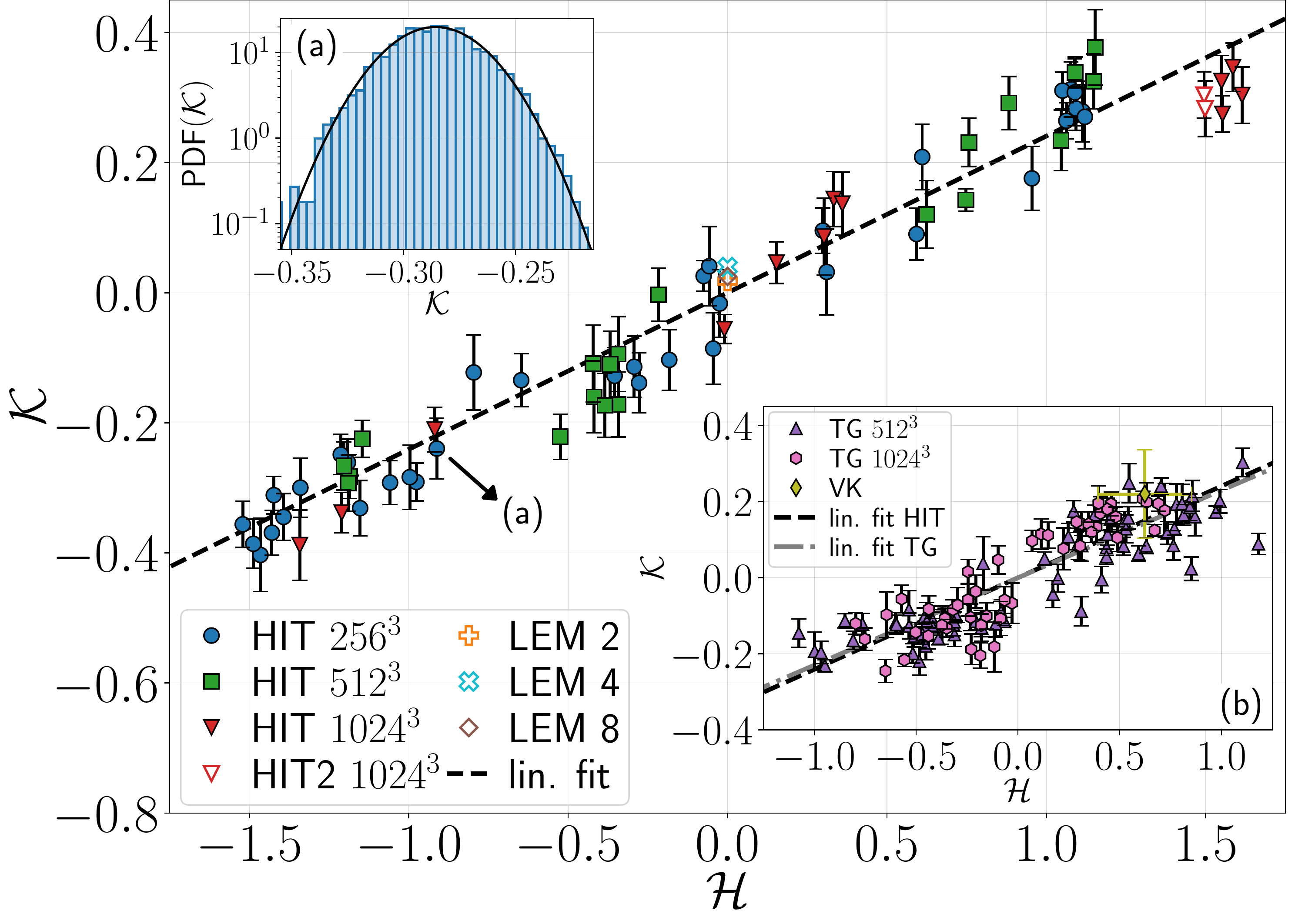}
    \caption{Linking number of tracers trajectories $\LN$ as a function of normalized helicity $\DH$ for DNSs of HIT and the LEM experiment. Error bars correspond to 95\% confidence intervals. A least-squares linear fit taking into account the error bars is shown as a reference.  Inset (a) shows the PDF of $\LN$ in semi-logarithmic scale for different subsets of tracers of the same DNS, indicated by the arrow. A Gaussian distribution with the same mean and standard deviation is shown for comparison.  Inset (b) shows $\LN$ as a function of $\DH$ for the VK experiment, and cells in TG simulations; each of the cells corresponds to a VK-like swirling flow with non-zero mean helicity. A weighted least-squares linear fit for all TG data is shown, and it is compared with the linear fit obtained from HIT.}
\label{fig:calibration}
\end{figure}

Figure \ref{fig:calibration} shows $\LN$ as a function of $\DH$ for DNSs of HIT and for three datasets of the LEM experiment. Error bars represent 95\% confidence intervals (approximately twice the standard deviation of $\LN$).  $\LN$ and $\DH$ were computed for $\Delta T = \tau_L$. For the entire range of $\DH$ explored, the data displays a linear dependence between the two quantities, irrespective of Reynolds number and flow geometry.  
Therefore, we propose that these two global, large-scale quantities are related by
\begin{equation} 
    \LN = \alpha~\DH,
\label{eq:result} 
\end{equation} 
where $\alpha$ is an unknown dimensionless constant. 
An additive constant is not expected in this relation,
as we can assume that a mirror-symmetric flow will have statistically as many $+1$ crossings as $-1$ crossings (which is consistent with the data). An error-weighted least-squares fit using the HIT data yields $\alpha_\text{HIT} = 0.241 \pm 0.006$ (95\% confidence level) independently of the Reynolds number (provided a fully developed turbulent state is reached). A dashed straight line with this slope is shown in Fig.~\ref{fig:calibration}. We verified that a linear relation as in Eq.~(\ref{eq:result}) holds regardless of the particular choice of $U$ and $L$ employed to normalize $H$ in Eq.~(\ref{eq:normalized_helicity}).

Before discussing the other datasets, note 
Eq.~(\ref{eq:result}) is robust: It holds for all datasets with small changes in $\alpha$ within errors, and when the number of particles $N$, the time span $\Delta T$, or other parameters are changed, and also 
when sufficiently large subregions of the flow are considered. 
To understand how using a finite number of trajectories affects the determination of $\LN$ and its error, different subsets of 250 particles were randomly chosen from the $10^6$ trajectories traced in each DNS. We see minor variations in the value of $\LN$, as shown by the probability distribution function (PDF) of $\LN$ for different subsets in a DNS (indicated by the arrow) in Fig.~\ref{fig:calibration}(a). The PDF is compatible with a Gaussian distribution and its dispersion is associated with the errors in $\LN$ when using a finite number of tracers. Note this PDF does not correspond to local helicity fluctuations: it is a measure of the uncertainty in the determination of $\LN$. 
Using $N = 250$ is motivated by the number of trajectories that can be simultaneously observed using PTV in experiments, of the order of the several tens to a few hundreds. Indeed, such value is enough to get a reasonable determination of $\LN$, with smaller errors as $N$ is increased.
As $\Delta T$ decreases, $N = 250$ is still enough to determine $\LN$, although with larger error bars.
By varying $\Delta T$, reasonable correlation (with the same value of $\alpha$) is obtained between $\LN$ and $\DH$ when $\Delta T \geq \tau_L/5$.  For $\Delta T < \tau_L/10$ uncertainties in $\LN$ prevent distinction between chiral and non-chiral flows. By increasing $N$ to $1000$, 
we verified that computing $\LN$ over $\Delta T = \tau_L/10$ allows for a distinction between different flow chiralities, but to achieve a correlation between $\LN$ and $\DH$ (consistent within uncertainties) it is still necessary to use $\Delta T \approx \tau_L/5$. This can be interpreted as a limit on how short a history of the flow is needed to reconstruct its topology.  Furthermore, the condition $\Delta T \geq \tau_L/5$ implies that if $\LN$ is used to estimate the helicity in a flow as a function of time, $\approx 0.2 \tau_L$ is the maximum time cadence for which $\LN(t)$ (and thus $\DH(t)$) can be estimated. Finally, increasing $\Delta T$ (for fixed $N$) results in a better agreement between $\LN /\alpha$ and $\DH$.  See \cite{SI} for more details on the uncertainty in the determination of $\LN$, and on the robustness of the results on $N$, $\Delta T$ and on the memory of the trajectories.

As previously mentioned, Fig.~\ref{fig:calibration} also shows data from tracers in LEM, an experiment
that generates mirror-symmetric (i.e., zero helicity) HIT. The turbulence generation mechanism is quite different from the DNSs, using multiple impellers instead of random volumetric forcing. Three experimental runs with different Reynolds numbers (labeled LEM 2, LEM 4 and LEM 8) were analyzed. $\LN$ was estimated from the signed crossings over $\Delta T = \tau_L$ for each run. 
The mean value of $\LN$ for each one is shown in Fig.~\ref{fig:calibration}; their values are $\LN_{\text{LEM~2}} = 0.018 \pm 0.18$, $\LN_{\text{LEM~4}} = 0.040 \pm 0.29$ and $\LN_{\text{LEM~8}} = 0.025 \pm 0.28$ (95\% confidence intervals). Such large fluctuations arise from time fluctuations in the flow, since the measurements were performed in the central region of the setup (of dimensions much larger than the Kolmogorov scale) where turbulence is expected to be more isotropic but large-scale fluctuations are likely to be strong as one impeller may  temporarily dominate over others.
Nonetheless, the value of $\LN$ is compatible with zero, consistent with null mean helicity.

{\it Helicity and the linking number of fluid trajectories in swirling flows.} The relation given by Eq.~(\ref{eq:result}) holds for other turbulent flows besides HIT, and even locally in space, provided the region is large compared to the Kolmogorov scale, as $\DH$ and $H$ are global, averaged quantities. We now consider the DNSs of TG turbulence and the VK laboratory experiments. Because of symmetries in the TG forcing \cite{Brachet_1983, Nore_1997} the flow in these DNSs can be divided into 8 cells each of volume $(\pi)^3$. The 8 cells are labeled as [$C_x$,$C_y$,$C_z$], where $C_i = 1$ or $2$, with $1$ labeling the region from $0$ to $\pi$ in the $i$-th direction, and $2$ the region between $\pi$ and $2\pi$ (i.e., the cell labeled [1,2,1] refers to the subregion $[0,\pi)\times[\pi,2\pi)\times[0,\pi)$ of the whole computational domain). Each cell has a flow that in many previous studies was shown to have Eulerian and Lagrangian similarities with that observed in VK experiments \cite{Ponty_2005, Mininni_2014, Angriman_2020}, despite differences in boundary conditions and forcing mechanisms (volumetric forcing in the former, and two counter-rotating propellers in the latter): two counter-rotating vortices separated by a shear layer. The VK flow has non-zero helicity, while given the TG symmetries, four of the DNS cells have mean helicity with a preferential sign, and the other four cells have the opposite sign.

\begin{figure} 
    \centering 
    \includegraphics[width=1\linewidth]{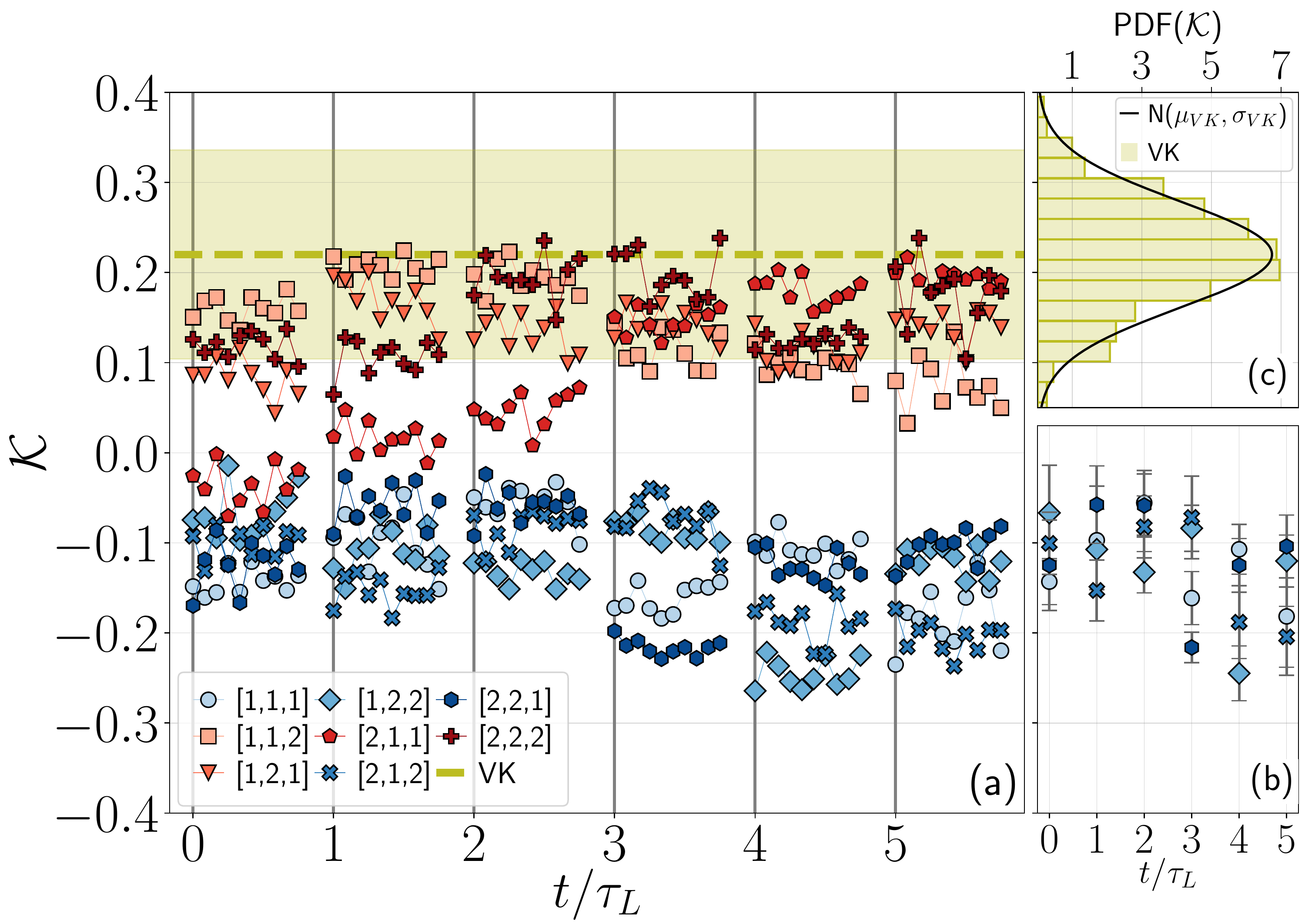}
    \caption{(a) Linking number $\LN$ as a function of time in the 8 TG cells, and in the VK experiment. The different points correspond to $\LN$ estimated for 10 subsets of $250$ particles (in groups of connected points), at different consecutive time intervals $\Delta T=\tau_L$ up to time $6 \tau_L$. Half the cells have $\LN>0$ (with fluctuations in time), and the other half $\LN<0$. The dashed line is the mean value of $\LN$ in the VK experiment, with the colored band representing a 95\% confidence interval considering time fluctuations. (b) Time evolution of $\LN$ for all the cells with negative helicity, reconstructed by averaging the 10 subsets from the data in (a). (c) PDF of $\LN$ fluctuations in the VK experiment, with a normal distribution $N(\mu_{VK},\sigma_{VK})$ with the same mean $\mu_{VK}$ and standard deviation $\sigma_{VK}$ for comparison.}
\label{fig:TG_vs_VK} 
\end{figure}

Figure \ref{fig:calibration}(b) shows $\LN$ for the VK experiment (with $\DH$ estimated from DNSs and large-scale flow geometry), and for each cell in the TG simulations, as a function of each $\DH$ value, for two different Reynolds numbers in $512^3$ or $1024^3$ TG DNSs. As before, in simulations $N=250$ and $\Delta T =  \tau_L$. Fluctuations of $\LN$ and $\DH$ are larger than in HIT as this flow can have wild variations of helicity with time. However, a linear relation between both quantities is again recovered. A weighted least-squares fit using both resolutions yields $\alpha_{\text{TG}} = 0.23 \pm 0.01$ (95\% confidence interval). A straight line with this slope is indicated as a reference in Fig.~\ref{fig:calibration}(b), as well as another with slope $\alpha_{\text{HIT}}$ for comparison; $\alpha_{\text{HIT}}$ and $\alpha_{\text{TG}}$ are compatible within error bars (thus, in the following $\alpha = \alpha_{\text{HIT}}$).
This also shows that the linking number of fluid trajectories in subregions of the flow (as the individual cells in the DNSs of TG) is proportional to the local flow helicity.
Figure \ref{fig:TG_vs_VK}(a) depicts the value of $\LN (t)$ for each TG cell in simulations with $1024^3$ grid points. For each cell, $\LN$ was computed with $\Delta T = \tau_L$ over 10 random subsets of trajectories (each with 250 trajectories).
The different points in each vertical stripe correspond to $\LN$ estimated for these 10 subsets (in groups of connected points for subsets in the same cells), at different consecutive time intervals up to time $6 \tau_L$. 
$\LN$ (and $H$) fluctuates strongly in time in each cell, 
but half the cells preferentially have $\LN>0$, and the other half $\LN<0$. Given a TG cell and a time interval $\Delta T$, by averaging over the 10 subsets a mean value of $\LN$ at said time interval is computed, and a time series $\LN(t)$ for each cell is thus reconstructed.
Figure~\ref{fig:TG_vs_VK}(b) shows the resulting time series of the linking number (or, except for the factor $1/\alpha$, the flow's normalized helicity) for the cells with negative helicity.

The mean value of $\LN$ in the VK experiment at a similar Reynolds number as the DNSs, obtained from PTV, is also shown in Fig.~\ref{fig:TG_vs_VK}(a) by the dashed line, with a shaded horizontal stripe indicating 95\% of the fluctuations. Figure \ref{fig:TG_vs_VK}(c) shows the PDF of $\LN$ in the VK experiment for 500 measurements (the vertical axis is shared by both panels), and a normal distribution with the same mean and standard deviation. The value of $\LN$ in the VK experiment and the TG cells with positive helicity are compatible within uncertainties, as expected from the similarities both flows share in their large-scale geometry,
and despite the differences in boundary conditions and forcing mechanisms. Thus, the flow helicity has an imprint in the number of crossings of particles, and for very different turbulent flows, with or without mean flows, and with different boundaries.

We showed that fluid elements tell a story of the topology of the underlying turbulent flow. It is known that the irreversibility of turbulence has an effect on trajectories \cite{Xu_2014}. Here, the flow topology affects particles by linking the trajectories. In a mirror symmetric flow, signed crossings average to zero. In a chiral flow that's not the case: Flows with positive helicity result in a positive average of signed trajectory crossings, while flows with negative helicity result in negative signed crossings. Moreover, when properly normalized these two quantities are linearly related, with a proportionality constant that appears independent of the Reynolds number, the boundary conditions, and the mechanism that generates the turbulence. 
The linking of the particle paths involves macroscopic length and time scales. This is a remarkable feature, as a particle might go through flow regions in which strong fluctuations could potentially erase the broken mirror symmetry of the flow, specially in the TG flow.
Statistical alignment between $\mathbf{u}$ and $\mathbf{\nabla} \times \mathbf{u}$, which may differ depending on flow helicity and was reported to take place preferentially in regions of low dissipation \cite{Pelz1985,moffatt_1985,Tsinober1985,Levich1987,Farge2001}, can only partially explain these observations. 
As the particles move following $\mathbf{u}$, they may also partially follow vorticity field lines in helical regions, but this can only happen in such specific regions. The relation between $\DH$ and $\LN$ for long times and large scales thus indicates a stronger, and non-trivial, impact of flow chirality in physical observables and in the memory of fluid particles.
The particle linking number thus defined connects a global topological invariant of the flow with a topological number of fluid trajectories. Moreover, the ratio of proportionality $\alpha$ is the same for very different flows, and in particular, for cases with helicity in the large-scale mean flow (i.e., TG and VK flows) as for HIT without a mean flow, both numerically and experimentally.
This suggests that $\alpha$ may characterize a universal property of turbulence. Such connection has implications, e.g., for studies of mixing. If particle's trajectories get more linked in a helical flow, then helicity can have an effect in mixing, something already
noted in studies of helical flows in biological systems \cite{Goldstein_2008,Meent_2010}. Finally, the relation between $\DH$ and $\LN$ provides a way to estimate helicity in laboratory experiments, a quantity which so far has eluded detailed laboratory characterization in turbulent flows. Indeed, one motivation to use small sets of particles or short trajectories, instead of the millions of long trajectories usually accessible in DNSs, was to probe the robustness of the particle linking number when used in conditions as those found in the laboratory.

\begin{acknowledgments}
This work was partially supported by the ECOS project A18ST04.
    S.A., P.J.C.~and P.D.M acknowledge support from grants PICT Nos.~2015-3530  and 2018-4298, and UBACyT No.~20020170100508. M.B., S.G.H.~and
    R.V.~acknowledge support from European Project EuHIT (European
High-Performance Infrastructures in Turbulence, grant No.~312778), and ANR-13-BS09-0009. Computational resources were provided by the HPC center DIRAC, funded by Instituto de Fisica de Buenos Aires (UBA-CONICET) and the SNCAD-MinCyT initiative. We gratefully acknowledge comments and suggestions by reviewers which helped improve our manuscript.
\end{acknowledgments}

\bibliography{ms}

\clearpage
\onecolumngrid
\section*{Supplemental Material: Broken mirror symmetry of tracer's trajectories in turbulence}

\section{Direct numerical simulations}
We performed direct numerical simulations (DNSs) of the incompressible Navier-Stokes equations
\begin{equation}
\frac{\partial \mathbf{u}}{\partial t} + \mathbf{u}\cdot\mathbf{\nabla}\mathbf{u} = - \mathbf{\nabla} p + \nu \nabla^2 \mathbf{u} + \mathbf{F},
\end{equation}
where $\mathbf{u}$ is the velocity field, $p$ the pressure per unit mass density $\rho_0$, $\nu$ the kinematic viscosity, and $\mathbf{F}$ an external volumetric forcing per unit mass density that sustains the turbulence. The equations are solved in a dimensionless three-dimensional $(2\pi)^3$-periodic cubic domain using a parallel pseudospectral method with the GHOST code \cite{Mininni_2011c, Rosenberg_2020}. For all forcing schemes considered, the flow is evolved until a turbulent steady state is reached, and then $10^6$ Lagrangian particles are injected in the flow, which evolve according to
\begin{equation}
\frac{d\mathbf{x}_p}{dt} = \mathbf{u}(\mathbf{x}_p, t),
\end{equation}
where $\mathbf{x}_p(t)$ is the position of the tracer at time $t$, and $\mathbf{u}(\mathbf{x}_p,t)$ is the velocity of the fluid at $\mathbf{x}_p(t)$. Integration of tracers is performed using a second-order Runge-Kutta time stepping scheme, and a three-dimensional third-order spline interpolation to estimate the fluid velocity $\mathbf{u}(\mathbf{x}_p,t)$ at the position of the particle. The instantaneous position and velocity of each particle is tracked as it evolves in time along with the fluid. Two different forcing schemes for $\mathbf{F}$ were considered.

\subsection{Homogeneous and isotropic turbulence with tunable helicity}

To generate homogeneous and isotropic turbulence (HIT) while controlling the amount of helicity injected in the flow, the mechanical forcing was chosen as a superposition of Fourier modes with random phases, and a correlation time of $0.5$ turnover times (random phases were slowly varied in time to prevent abrupt changes in the forcing). Helicity injection was controlled using the method introduced in \cite{Pouquet_1978}: Two random independent fields $\mathbf{q}^{(1)}$ and $\mathbf{q}^{(2)}$ are generated in Fourier space, each normally distributed and centered around $k_F$, the forcing wave number. From these, two normalized and incompressible fields $\mathbf{f}^{(1)}$ and $\mathbf{f}^{(2)}$ are defined as
\begin{equation}
\mathbf{f}^{(i)} = \frac{\mathbf{\nabla}\times \mathbf{q}^{(i)} }{ \mathbf{\langle |\nabla}\times \mathbf{q}^{(i)}|^2\rangle^{1/2} }, \quad i=1,2.
\end{equation}
Lastly, by correlating the fields $\mathbf{f}^{(1)}$ and $\mathbf{f}^{(2)}$, the mechanical forcing in Fourier space is given by
\begin{equation}
\mathbf{F}(\mathbf{k}) = f_0 \Big\{\text{cos}(\zeta) \mathbf{f}^{(1)}(\mathbf{k}) + \text{sin}(\zeta) \mathbf{f}^{(2)}(\mathbf{k}) + \frac{1}{k} \mathbf{\nabla}\times  [ \text{sin}(\zeta) \mathbf{f}^{(1)}(\mathbf{k}) + \text{cos}(\zeta) \mathbf{f}^{(2)}(\mathbf{k})]  \Big\},
\end{equation} 
where $f_0$ is the amplitude of the forcing, and $\zeta$ controls the amount of helicity injected in the flow. The relative helicity of the forcing is given by $\sin(2\zeta)$, so $\zeta = 0$ corresponds to no helicity injection (on average), while $\zeta = \pi/4$ results in maximal helicity injection. For the HIT DNSs, three Reynolds numbers were considered respectively with spatial resolutions of $256^3$, $512^3$ and $1024^3$ grid points. For each Reynolds number and spatial resolution, multiple runs with $\zeta \in [-\pi/4,\pi/4]$ were done (i.e., varying the amount of helicity in the flow). Also, a HIT simulation (``HIT 2'') with a forcing correlation time of $1/100$ of the turnover time was performed, with a spatial resolution of $1024^3$ grid points and with $\zeta = \pi/4$, to study whether changing other forcing parameters, as the forcing correlation time, had a significant effect in the observed correlation between helicity and the tracers' linking number.

\subsection{Taylor-Green DNSs}
For these simulations the external mechanical forcing is based on the Taylor-Green flow \cite{Brachet_1983}:
\begin{equation}
F_x = F_0~ \text{sin}(k_F x)~\text{cos}(k_F y)~ \text{cos}(k_F z), 
  \,\,\,\,\, 
F_y = - F_0~ \text{cos}(k_F x)~\text{sin}(k_F y)~ \text{cos}(k_F z),
  \,\,\,\,\, 
F_z = 0,
\end{equation}
with forcing wave number $k_F = 1$. The resulting flow presents several symmetries in a statistical sense, see Ref.~\cite{Brachet_1983, Nore_1997}, and as a result the flow in the full domain can be split into eight cells of volume $\pi^3$. In each cell the flow consists of two counter-rotating large-scale vortices which lie perpendicular to $\hat{z}$, separated by a shear layer in the mid plane. The flow in each cell has a similar geometry to the von K\'arm\'an experiment (see Sec.~2\ref{sec:VK}), both from a Eulerian and a Lagrangian point of view \cite{Mininni_2014, Angriman_2020}. Also because of the symmetries the total helicity in the $(2\pi)^3$ domain is zero, but each $\pi^3$ cell has non-zero helicity which fluctuates in time (the mean value alternates between positive and negative values when neighboring cells are crossed). With this forcing two different Reynolds numbers were explored, using spatial resolutions of $512^3$ and $1024^3$ grid points.

\begin{figure}[t!]
\centering
    \includegraphics[width=1\linewidth]{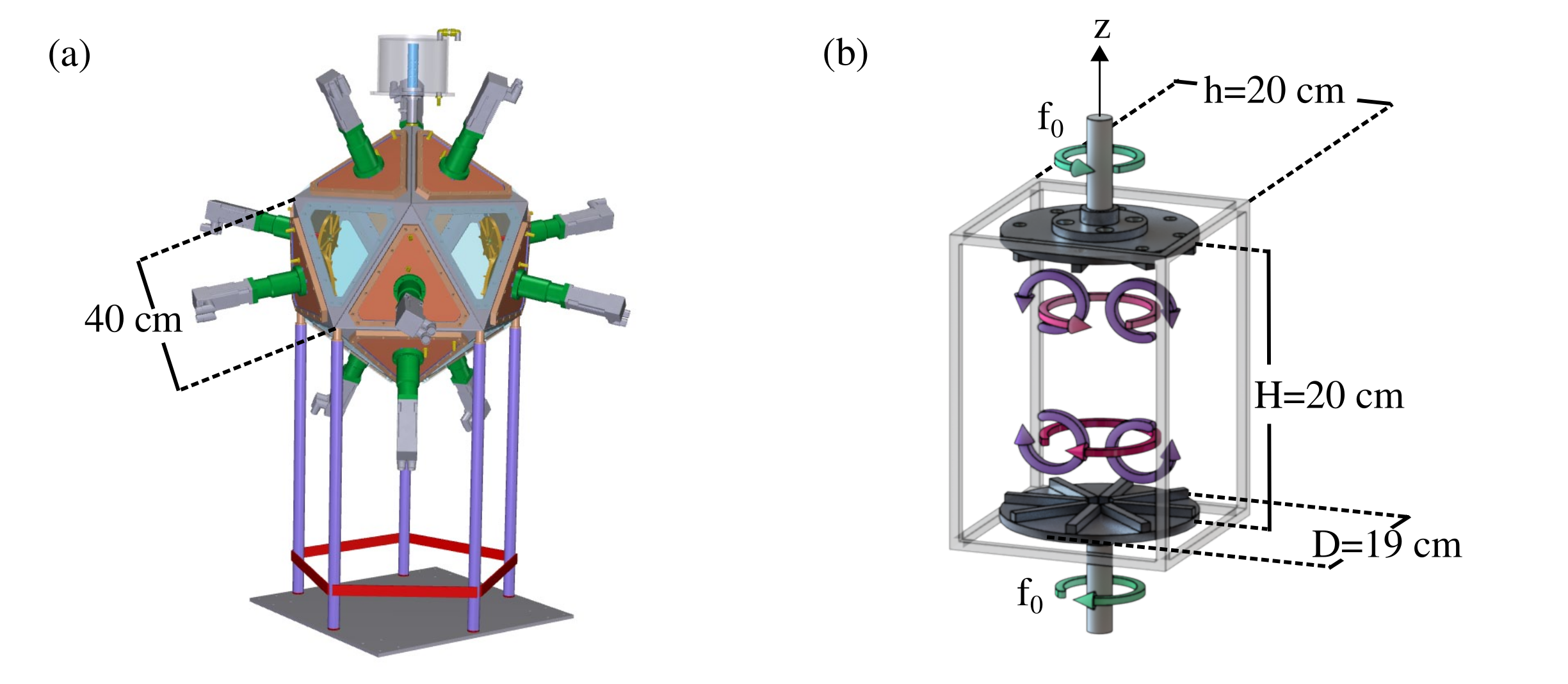}
    \caption{(a) Schematic view of the LEM setup. Twelve impellers are positioned on twelve of the twenty faces of the vessel containing the fluid. The length of the size of each triangular face is indicated as a reference, as well as the position of the motors and the windows used for measurements. (b) VK setup, with a representation of the mean large-scale flow. Two propellers with straight blades are used; the size of the vessel and the distance between propellers are indicated as a reference.}
    \label{fig:exp_setups}
\end{figure}

\section{Laboratory experiments}
\subsection{The Lagrangian Exploration Module (LEM)}
The experimental data for HIT was obtained from the LEM setup in Lyon \cite{LEM}. The setup consists of twelve independently controlled impellers with a diameter of $10$~cm, located at twelve of the twenty faces of a regular icosahedron and placed in a configuration to ensure maximum isotropy and mirror symmetry. The edges of the icosahedron have a length of $40$~cm, so that the full volume of the vessel is $\approx 140$~l. The vessel is filled with degassed and filtered water. A schematic view of the setup is depicted in Fig.~\ref{fig:exp_setups}(a). Observation windows are made of polymethyl methacrylate (PMMA). Each impeller is driven by an independent brushless motor (Unimotor, Leroy Somer, $640$~W). For the experiments in this work the forcing is isotropic: all impellers are used with the same constant rotation frequency $f^{\text{LEM}}_{0}$, achieving a turbulent steady state in all cases. The turbulent flow at the center of the apparatus---in a region comparable to the flow integral scale---is statistically homogeneous and isotropic, with a mean flow whose magnitude is about 10\% of the turbulent fluctuating velocity and has zero mean helicity.

Polyethylene microspheres with diameters of $106$--$125~\mu$m (Cospheric) are used as tracers (density $0.995$~g~cm~$^{-3}$). The setup is illuminated using a pulsed Nd:YAG laser (Quantronix Dual Condor, with mean output power of $150$~W and with repetition rates synched with the cameras) with wavelength of $532$~nm and pulse width of $200$~ns. The laser is collimated and expanded to illuminate the measurement volume. Three high-speed cameras (Phantom V12, Vision Research Inc.) are used to measure the flow in a volume of $5\times 6.5\times 5.5$~cm$^3$, with an imaging resolution of $50~\mu$m/px. The optical system is calibrated to recover the instantaneous 3D position of each particle, which is then tracked in time \cite{Bourgoin_2020}. The velocity for each track is derived from the trajectories using Particle Tracking Velocimetry (PTV).

Several experimental runs were made varying the rotation frequency of the impellers, $f_{0}^{\text{LEM}} \in \{2, 4, 8\}$~Hz. The cameras and laser's frame rate was chosen at least one order of magnitude faster than the Kolmogorov time scale $\tau_\eta$. 
There are approximately $210$ different trajectories in time intervals of duration $\tau_L$ (the Lagrangian velocity correlation time, defined below), with a mean duration of $0.08 f_0^{\text{LEM}}$. After many runs for each forcing frequency, there are a total of $\mathcal{O}(10^{5})$ 3D trajectories available.

\subsection{The von K\'arm\'an Swirling Flow experiment}\label{sec:VK}
The data come from a von K\'arm\'an (VK) flow experiment in Buenos Aires \cite{Angriman_2020}. The setup comprises two facing disks of diameter $D = 19$~cm, separated by a vertical distance of $H = 20$~cm, each fitted with eight straight blades. The blades have a height of $1$~cm, a width of $1$~cm and a length such that they do not reach the center of the disk. The propellers are contained in a PMMA cell of square cross-section with side $h = 20$~cm, giving access to an experimental volume of ($20 \times 20 \times 20$)~cm$^3$ where the flow can be measured. The total size of the cell is ($20 \times 20 \times 50$)~cm$^3$, leaving space in the back of the propellers for shafts that connect the propellers to motors, and for refrigeration coils that allow heat removal if needed. Each propeller is driven by an independent brushless rotary servomotor (Yaskawa SMGV-20D3A61, 1.8 kW) controlled by a servo controller (Yaskawa SGDV-8R4D01A) which provides access to the instantaneous velocity and torque of the motor.
The cell is filled with distilled water from a double pass reverse osmosis system, to remove ions and suspended solid particles. For this study the two disks rotate in opposite directions at an angular rotation frequency $f^{\text{VK}}_0$, stirring the fluid in the cell. This generates two large counter-rotating circulation cells producing, on average, a strong shear layer in the midplane between the disks. A secondary circulation in the axial direction is also generated by the propellers, resulting in a fully three-dimensional turbulent flow. A schematic representation of the VK setup and its mean flow is shown in Fig.~\ref{fig:exp_setups}(b).
The macroscopic flow structure and lack of mirror symmetry implies that the VK flow has non-zero helicity.

For each individual experimental run the flow is stirred by setting the angular frequency $f^{\text{VK}}_0 = 1.25$~Hz, so that the Reynolds number attained is comparable to the Reynolds number of the TG DNS using $1024^3$ grid points. The flow is seeded with tracer particles, which are neutrally buoyant polyethylene microspheres (density of 1 g~cm$^{-3}$) of diameter $d = 250$--$300~\mu$m (Cospheric). Particles were coated with a biocompatible surfactant (Tween 80) to ensure proper placement in suspension.

\begin{table*}[t!]
\begin{tabular*}{\textwidth}{l @{\extracolsep{\fill}} r c c c c c c c c c}
\hfill
Flow & Label & Datasets & $u$ & $\tau_\eta$ & $\nu$ & $\tau_L/\tau_\eta$  & $\varepsilon$ & $\tau_L/dt_{\text{p}}$ & $Re_{\text{part}}$& $Re_\lambda$\\
& & & [m/s] & [$10^{-3}$ s] & [$10^{-6}$ m$^2$/s] &  & [$10^{-2}$ m$^2$/s$^3$] &   \\
\hline
HIT & $256^3$ & 9 & 0.53 & 147 & 1200 & $14.3$ & $5.59$ & 140 & $490$ & 133 \\
 & $512^3$ & 5 & 0.57 & 102 & 480 & $22.0$ & $4.60$ & 450 & $1500$& 270  \\
 & $1024^3$ & 3 & 0.63 & 54 & 200 & $42.5$ & $6.84$ & 920 & $4600$ & 420  \\
\hline
HIT2 & $1024^3$ & 1 & 0.68 & 54 & 200 & $47.2$ & $6.95$ & 1010 & $5770$ & 475  \\
\hline
TG & $512^3$ & - & 0.84 & 53 & 675 & $40.5$ & $24$ & 430 & $2300$ & 215  \\
& $1024^3$ & - & 0.85 & 36 & 300 & $48.5$ & $23$ & 700 & $4200$ & 337  \\
\hline
LEM & LEM 2 & - & $0.062$ & 12  & 1 & $15.6 $& $0.66$ & 585 & $730$ & 180\\
    & LEM 4 & - & $0.118$ & 4.2 & 1 & $19.7$ & $5.5$  & 260 & $1200$ & 230\\
    & LEM 8 & - & $0.241$ & 1.5 & 1 & $28.9$ & $45$ & 270 & $2500$ & 336\\
\hline
VK  & - & - & 0.17 & 3.8 & 1 & $56.1$ & $6.6$ & 109 & $6300$ & 435 \\    
\end{tabular*}
% \end{ruledtabular}
\caption{Parameters of all datasets. DNS data is dimensionalized using a unit length $L_0=1$ m, a unit velocity $U_0= 1$ m$/$s, and a unit density $\rho_0 = 1$ kg$/$m$^3$. For the DNSs, the label corresponds to the grid resolution. For the LEM experiment, labels indicate the impellers frequency in Hz. For the HIT simulations the number of datasets (``Datasets'') is equal to the total number of simulations with different values of $\zeta$ (also, each simulation was conducted for several turnover times, allowing multiple estimations of the helicity). ``HIT 2'' corresponds to a HIT run with a very short forcing correlation time. $u$ is a measure of the 1D r.m.s.~value of the particles' velocity, $\tau_\eta = (\nu/\varepsilon)^{1/2}$ is the Kolmogorov time scale with $\nu$ the kinematic viscosity, and $\varepsilon$ the energy injection rate. $\tau_L$ is the Lagrangian correlation time, $dt_p$ is the particles' sampling time, $Re_\textrm{part}$ is the Reynolds number based on the particles' velocities, and $Re_\lambda$ the Taylor-based Reynolds number.}
\label{table1}
\end{table*}

Measurements of particles' dynamics were carried out using PTV. The cell is illuminated from two adjacent sides using two ($25\times 25$)~cm$^{2}$ LED panels (each 1880~lm, 22~W). Two high-speed cameras (Photron FASTCAM SA3) capture the particle's shadow projection over a bright background on two perpendicular sides of the cell. The cameras are aligned in such a way that the center of each image coincides with the center of the face of the cell that is being recorded. One camera captures the $x$--$z$ components of the particle's position, while the other measures the $y$--$z$ components, so the 3D individual trajectories are later reconstructed from 2D trajectories in each view. Each camera has a maximum speed of $1000$~fps at full resolution of $(1024\times 1024)~\text{px}^2$, and 12-bit color depth.
The cameras are placed in front of the cell at a distance $L = 3.5$~m so that the region of observation of $(16\times 16\times 16)~\text{cm}^{3}$ covers nearly the whole experimental volume, while warranting minimal optical distortion, with a spatial resolution of $0.16$~mm/px. 
We employ a $70$-$300$~mm lens in each camera, using a focal length of $260$~mm. Under these experimental conditions, the estimated maximum error in imaging a particle's position due to perspective effects is $\approx 325~\mu$m, which is of the order of the tracers' diameter (see Ref. \cite{Angriman_2020} for more details). For the results presented here, the sampling frequency of the cameras is set at $f_s \approx 1/\tau_{\eta}$. After several realizations of the experiment there are $\mathcal{O}(10^4)$ 3D trajectories with a mean duration per trajectory of $0.34/f_0^{\text{VK}}$. In each time interval of duration $\tau_L$ there are approximately $105$ trajectories available. From the individual trajectories, the instantaneous velocity is derived after applying a Gaussian filter.

\begin{figure}[t!]
    \centering
  \includegraphics[width=.49\linewidth]{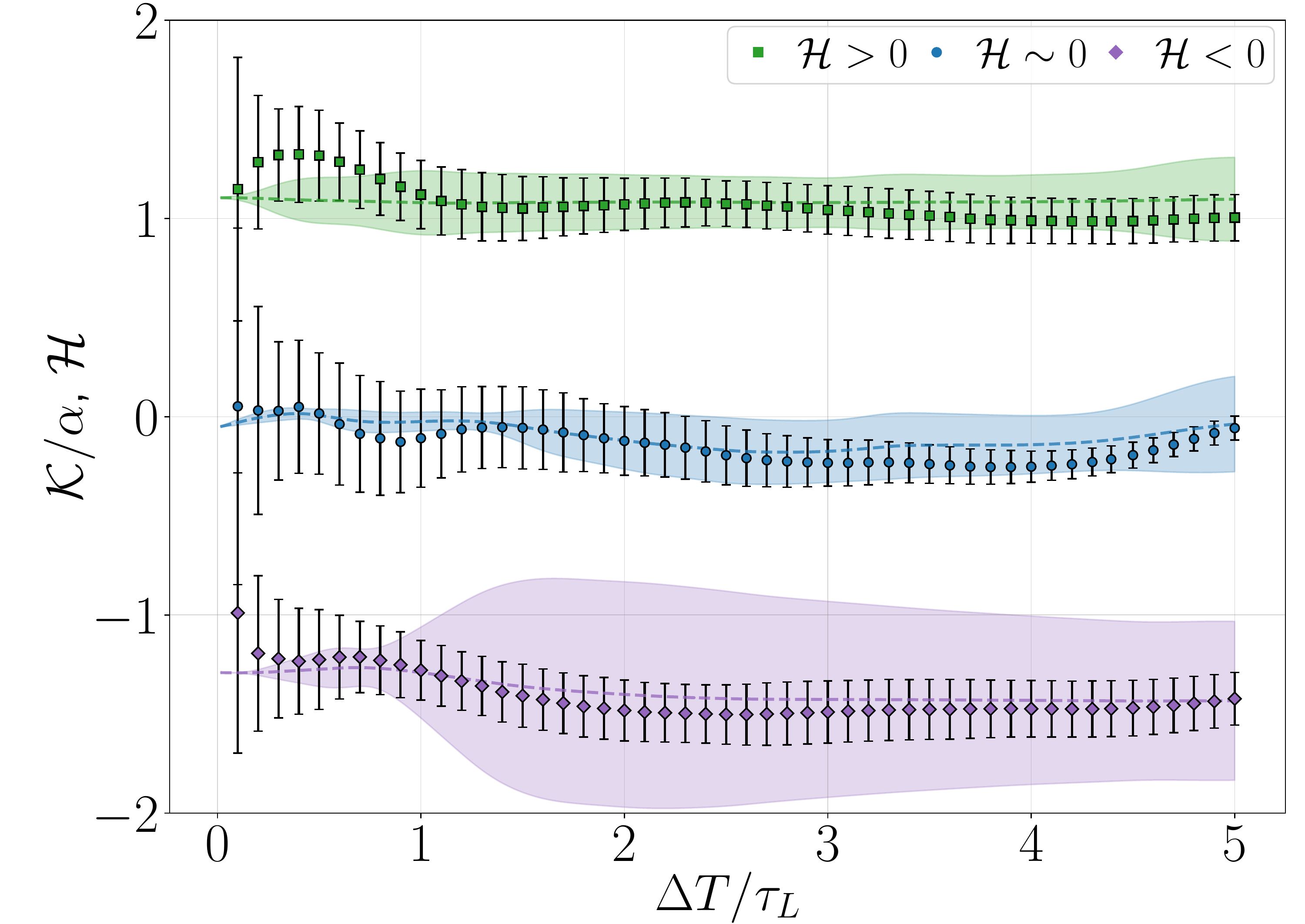}
  \includegraphics[width=.49\linewidth]{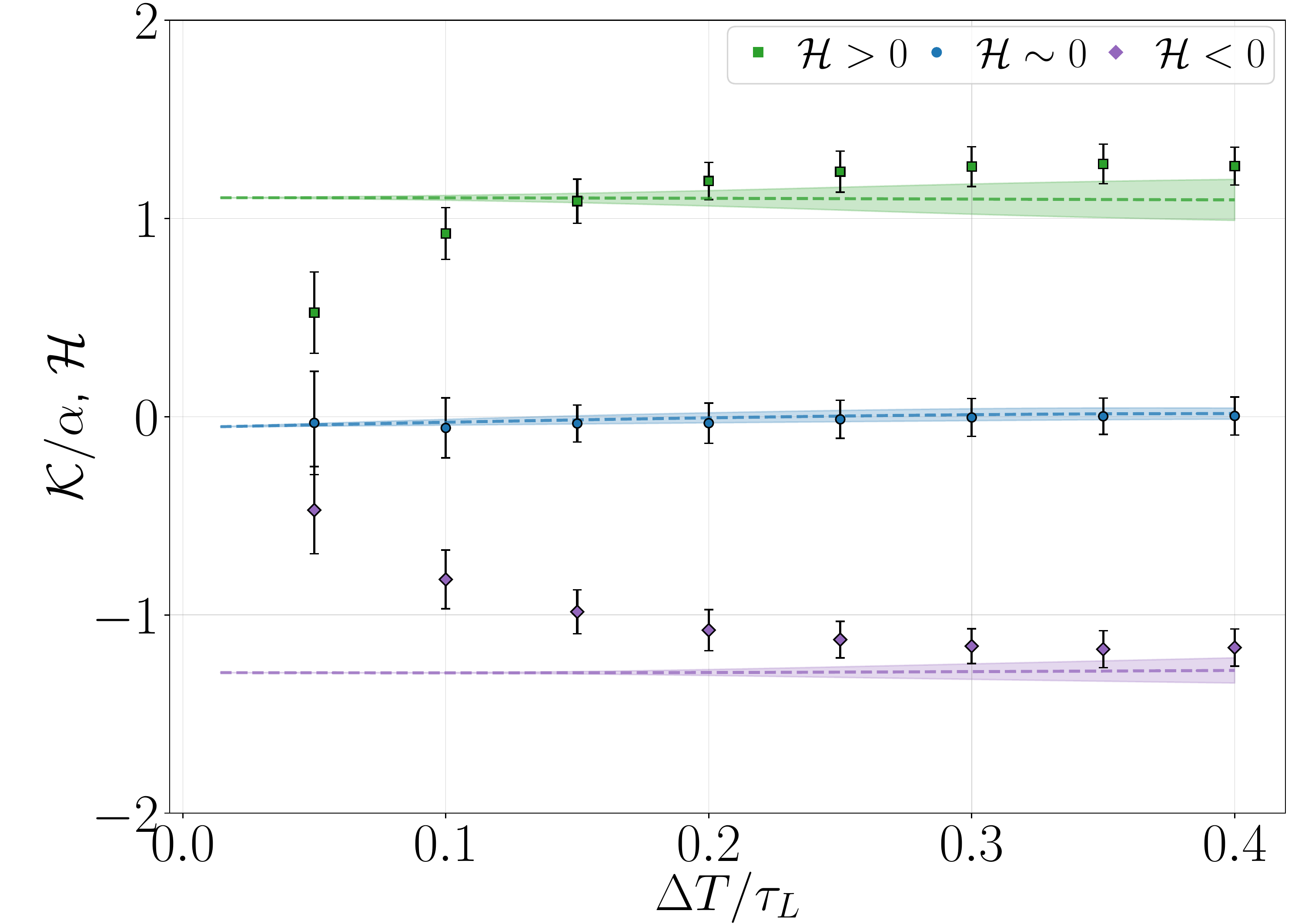}
    \caption{{\it Left:} Estimation of the helicity from the particles' linking number for different time spans $\Delta T$ (in units of $\tau_L$), using subsets of 250 particles. The dashed lines represent $\altmathcal{H}$ averaged over $\Delta T$ for three simulations, and the shaded regions indicate their corresponding standard deviation (associated to the strength of helicity fluctuations in time). The markers with error bars depict $\altmathcal{K}/\alpha$ (i.e., and estimation of the flow helicity) computed over a given time span $\Delta T$.
      {\it Right:} Estimation of $\altmathcal{K}/\alpha$ and $\altmathcal{H}$ using subsets of $1000$ trajectories, as a function of $\Delta T$. Labels are the same as in the left panel.}
    \label{fig:min_sampleo}
\end{figure}

\section{Flow characterization}

To define Reynolds numbers and characterize the flows we use the one-component r.m.s.~tracers' velocity $u$. For isotropic flows (HIT DNSs and the LEM experiment) $u$ is defined as $u = (U/3) ^{1/2} = (\langle v_x^2 + v_y^2 + v_z^2\rangle /3)^{1/2}$, where $v_i$ is the $i^{\text{th}}$ component of the tracer's velocity and the brackets $\langle \cdot \rangle$ denote averages over time and over all trajectories. For anisotropic flows (TG and VK), $u$ is computed using the horizontal components of the velocity (i.e., $v_x$ and $v_y$): $u = (\langle v_x^2 + v_y^2 \rangle /2)^{1/2}$. Note that to normalize the kinetic helicity we always use $U = (\langle v_x^2 + v_y^2 + v_z^2\rangle)^{1/2}$, for all flows, motivated by the fact that the helicity is a volumetric and three-dimensional quantity.

The Lagrangian velocity autocorrelation function is computed for each component of the tracers' velocity as
\begin{equation}
R_L^{(i)}(\tau) \equiv \frac{C_v(\tau)}{C_v(0)} = \frac{\langle v_i(t)~v_i(t+\tau) \rangle}{\langle v_i^2(t) \rangle},
\end{equation}
with $\tau$ the time lag. The one-dimensional Lagrangian correlation time $\tau^{(i)}_L$ is estimated from this correlation function. For isotropic flows the correlation time is then defined as $\tau_L = (\tau_L^{(x)} + \tau_L^{(y)} + \tau_L^{(z)})/3$, whereas for the TG and VK flows $\tau_L = (\tau_L^{(x)} + \tau_L^{(y)})/2$. A different procedure was followed in the LEM experiments where the available trajectories are short, as explained below.

In the simulations the energy injection rate $\varepsilon$ is readily available. In the VK experiments $\varepsilon$ is estimated using $u$ and $\tau_L$ as
\begin{equation}
\varepsilon = \frac{2}{C_0} \frac{u^2}{\tau_L},
\end{equation}
where $C_0 \approx 4$ is the Lagrangian second order structure function constant for the VK flow at the Reynolds numbers explored here (see \citep{Angriman_2020}, note $C_0$ can typically vary between 2 and 7 \citep{Sawford_2011}). For the LEM experiments, the energy injection rate is estimated from the Eulerian second order structure function. Note that in this case, given the shorter particle trajectories, the Lagrangian correlation time is then estimated as $\tau_L = 2 u^2/(\varepsilon C_0)$, with $C_0 \approx 6.0$ the constant for HIT at the experiments' Reynolds number. We verified that in the DNSs and VK experiment this procedure resulted in compatible estimations of $\tau_L$ within uncertainties. An integral Reynolds number based on tracers' velocity $Re_{\text{part}}$ can be defined in all cases as
\begin{equation}
Re_{\text{part}} = \frac{u L}{\nu},
\end{equation}
where $L$ is a characteristic length scale based on $u$ and $\tau_L$,
\begin{equation}
L = u \tau_L.
\end{equation}
Note that as $L$ is based on Lagrangian measurements, for it to be properly estimated tracers must be tracked for a sufficiently long time.
Using $u$ and the corresponding estimation of $\varepsilon$, the Taylor-based Reynolds number is defined as
\begin{equation}
Re_{\lambda} = \sqrt{\frac{15 u^4}{\nu \varepsilon}}.
\end{equation}
All the relevant parameters for the datasets analyzed are provided in Table \ref{table1}.

\begin{figure}[t!]
    \centering
  \includegraphics[width=.49\linewidth]{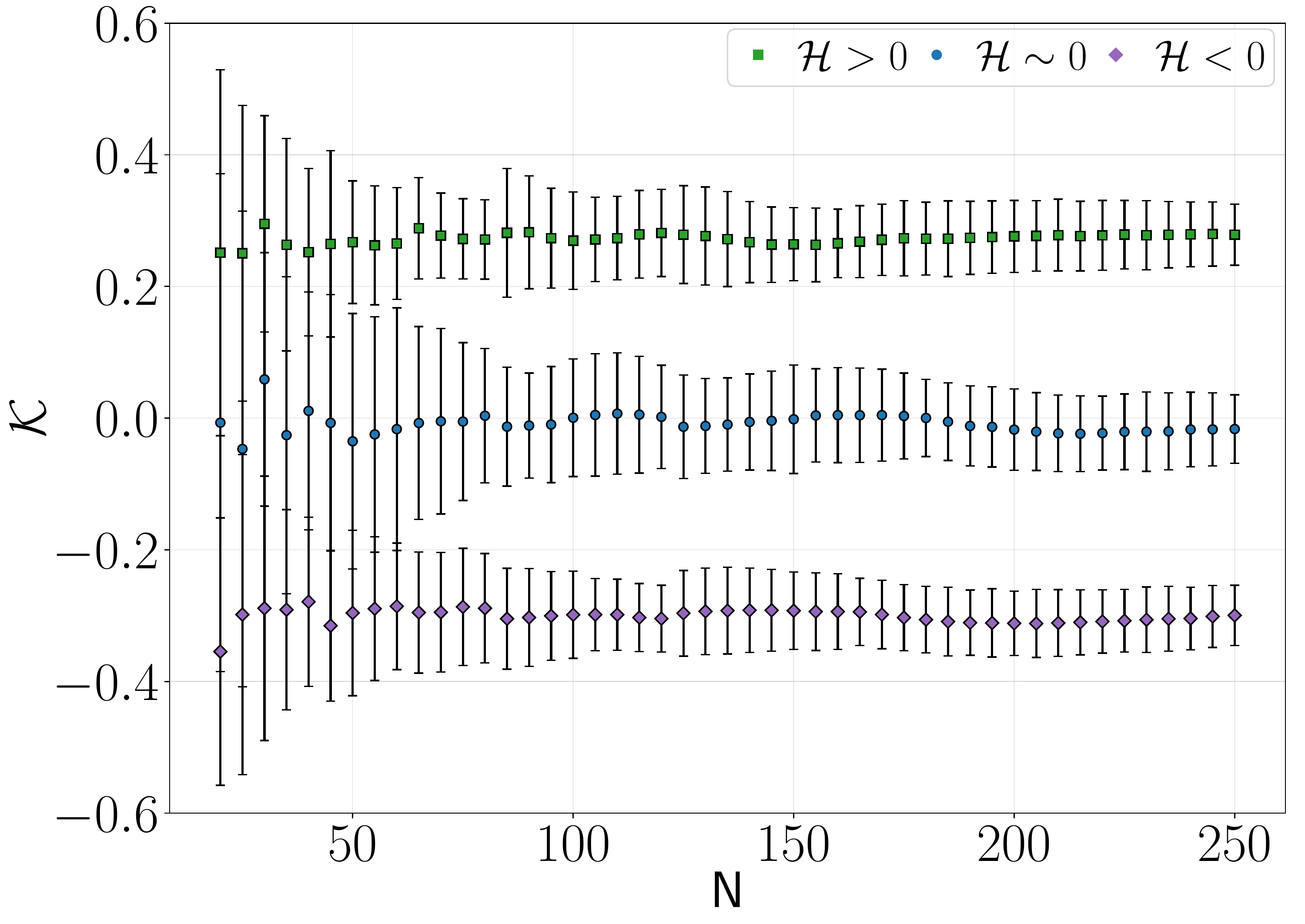}
  \includegraphics[width=.49\linewidth]{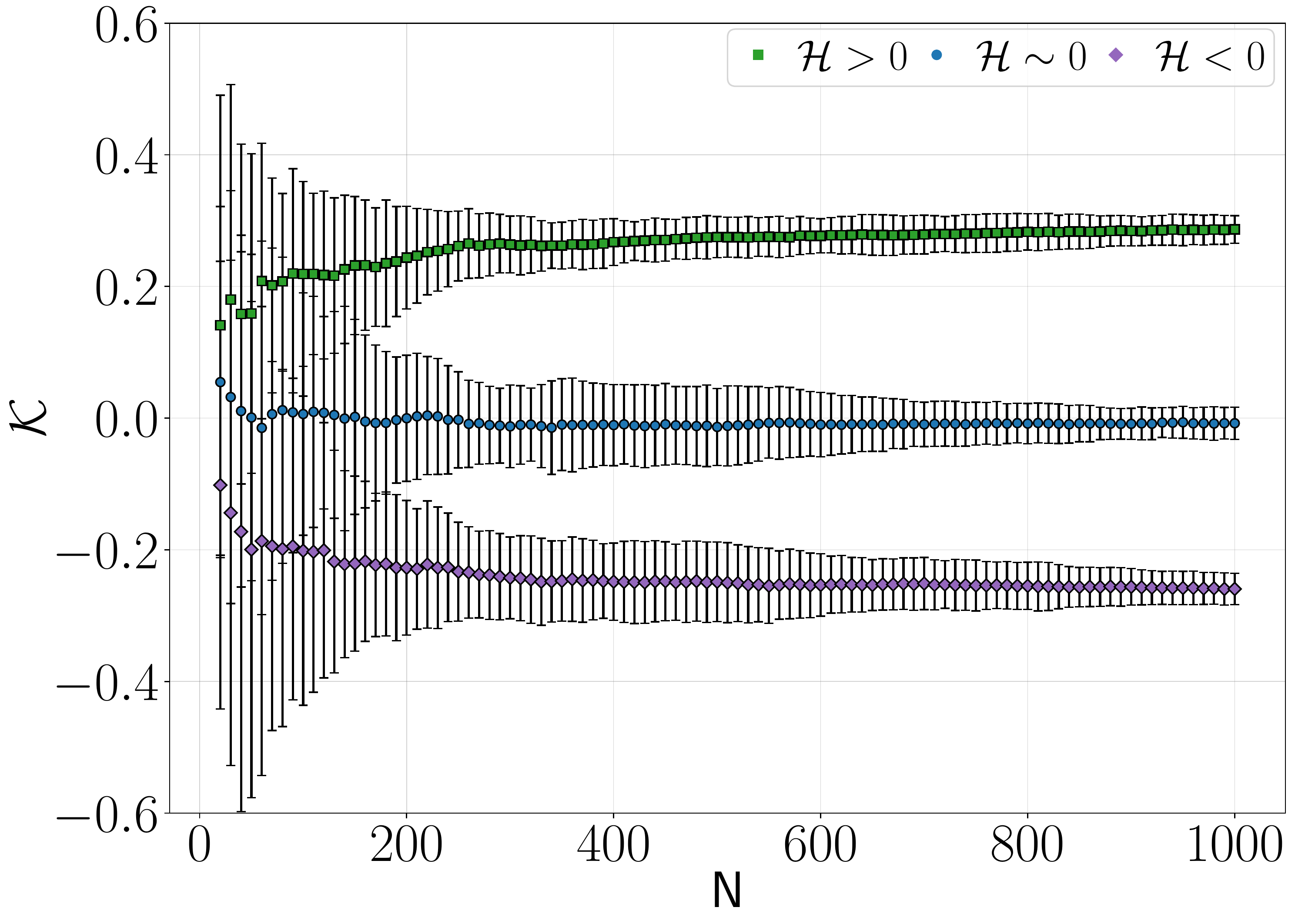}
    \caption{
	 Particles' linking number $\altmathcal{K}$ as a function of the number of trajectories used to determine the crossings, for three simulations over a fixed time interval $\Delta T$. Error bars in $\altmathcal{K}$ correspond in all cases to 95\% confidence intervals. {\it Left: $\Delta T = \tau_L$}   
    {\it Right:}  $\Delta T = \tau_L/5$.
    }
    \label{fig:K_Nparts}
\end{figure}

\section{Robustness of the results}

To analyze the robustness of the results we consider data from DNSs of HIT with $256^3$ grid points for three flow configurations: one with $\altmathcal{H} < 0$, one with  $\altmathcal{H} \approx 0$, and one with $\altmathcal{H}>0$. Similar results were obtained at other Reynolds numbers.

\subsection{Time interval dependence}

We first consider what is the minimum time span $\Delta T$ needed to estimate the particles' linking number such that flows with different helicity can be distinguished. For the three simulations mentioned above, the normalized crossings were computed over increasing time intervals $\Delta T$ from $\tau_L/10$ to $5\tau_L$, averaging $\LN$ over 20 sets of 250 randomly picked particles in each set. The quotient $\LN/\alpha$ (with 95\% confidence intervals) is shown in Fig.~\ref{fig:min_sampleo} as a function of $\Delta T$, and compared against the normalized helicity $\DH$ in the flow averaged over the same time span (with error bars corresponding to the standard deviation in the time fluctuations of the helicity). For $\Delta T = \tau_L/10$ the 95\% confidence intervals do not allow for statistical discrimination between the three cases with $\altmathcal{H} < 0$, $\approx 0$, and $>0$. However, just computing the crossings over $\Delta T = \tau_L/5$ is enough to distinguish between chiral and non-chiral flow states, and as $\Delta T/\tau_L$ increases the agreement between $\LN/\alpha$ and $\DH$ improves. Note also that the shapes of the curves $\LN(\Delta T)/\alpha$ follow qualitatively those of $\DH(\Delta T)$ for every dataset.

To see whether this is the smallest time interval at which helicity can be obtained from the particle linking number, we consider larger subsets of 1000 randomly picked particles. Figure \ref{fig:min_sampleo} (right) shows $\LN$ in this case for $\Delta T \leq 0.4 \tau_L$. In this case the different flow states are distinguishable for $\Delta T \geq 0.1 \tau_L$, as error bars are significantly reduced, but in order to consistently link $\LN$ to the value of helicity $\DH$ (within statistical uncertainties), $\Delta T$ needs to be, again, at least $0.2 \tau_L$. This indicates a minimum history of the trajectories is needed to reconstruct the flow topology, at least for the number of particles considered here.

\subsection{Number of particles}

While DNSs with millions of particles are feasible, experiments can often track only a few hundreds of particles at a time, although measurements can be repeated multiple times. For practical purposes, the determination of $\altmathcal{K}$ should be robust to the number of particle trajectories $N$ considered in each measurement. For a fixed time span $\Delta T = \tau_L$, we computed $\altmathcal{K}$ averaging over the crossings in 20 sets of $N$ particles (randomly picked for each set, from the total of $10^6$ particles available in each DNS). The number of particles was increased successively from $N = 20$ up to $N = 250$, in a range motivated by the typical amount of simultaneously available particle trajectories in a typical PTV laboratory measurement. The value of $\altmathcal{K}$ as a function of $N$ is shown in the left panel of Fig.~\ref{fig:K_Nparts} with 95\% confidence intervals, again for the three reference simulations with positive, negative, and zero helicity. For $N \geq 65$ it is possible to statistically discern between the different flow chiral states. The mean value of $\altmathcal{K}$ does not show significant variations, and as expected the error bars decrease with increasing $N$. 
As discussed before, if the time interval $\Delta T$ is decreased, error bars and uncertainties increase. Or, otherwise, a larger number of trajectories is needed for the statistics to converge with the same uncertainties. The right panel of Fig.~\ref{fig:K_Nparts} shows $\altmathcal{K}$ as function of $N$ but for a time interval $\Delta T = \tau_L/5$. In this case, $N\geq 190$ is enough to identify each flow state within statistical uncertainties, and larger values of $N$ provide better results. Moreover, the determination of $\LN$ using $\tau_L/5$ is compatible within error bars to that obtained for $\Delta T = \tau_L$.

\begin{figure}[t!]
    \centering
  \includegraphics[width=.49\linewidth]{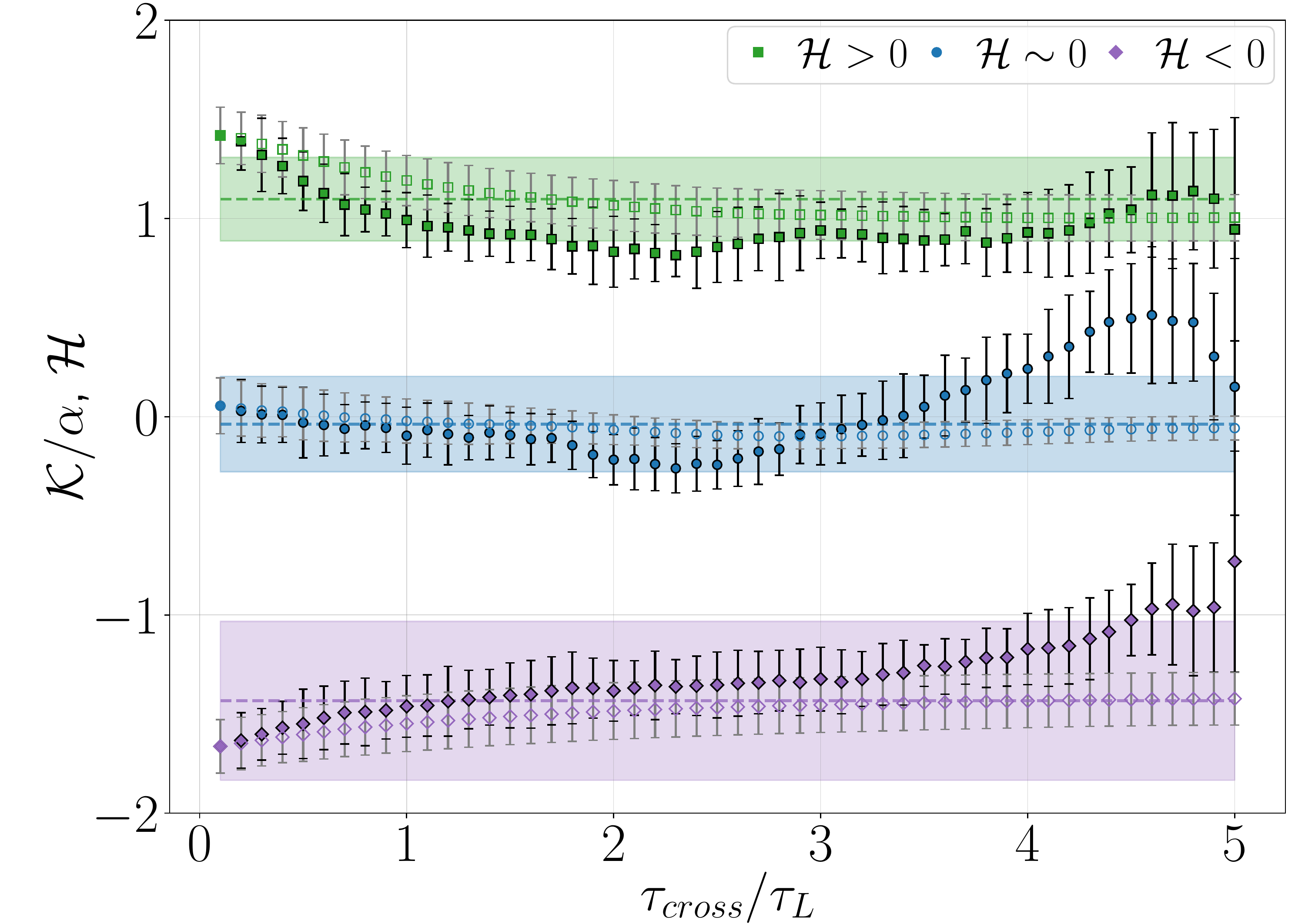}
    \caption{Normalized particles linking number $\altmathcal{K}/\alpha$, conditioned to crossings of trajectories occurring only in a window of time centered around $\tau_{cross}$ (in units of $\tau_L$, solid markers). The empty markers show the accumulated value of $\altmathcal{K}/\alpha$ for all crossings separated in time up to a time interval $\tau_{cross}$. All error bars correspond to 95\% confidence intervals. As a reference, the horizontal dashed lines with shaded regions indicate the mean flow helicity and its standard deviation.}
    \label{fig:K_cond}
\end{figure}

\subsection{Memory of the crossings}

Are all crossings of trajectories relevant? Or are crossings more important for the determination of $\altmathcal{K}$ if the particles were nearby at the time of the crossing, or on the contrary, very far away? For a time span $\Delta T = 5\tau_L$, and using 20 sets of 250 particles, we computed $\altmathcal{K}$ conditioned to only crossings occurring separated by a time interval close to $\tau_{cross}$. That is to say, given a crossing for which one particle passed through the crossing at time $t_0$, and the other particle passed through it at a time $t_0 + \delta \tau$, we only computed the crossing if $\delta \tau \in [\tau_{cross} - \tau_L/10, \tau_{cross})$. The result is shown in the right panel of Fig.~\ref{fig:K_cond}. Interestingly, consistent results are obtained for a very long range with $\tau_{cross} \lesssim 5 \tau_L$, i.e., even crossings separated in time contain information on the flow chirality. The figure also shows $\altmathcal{K}$ accumulated for all time separations up to $\tau_{cross}$, i.e., counting all crossings with $\delta\tau \in [0, \tau_{cross})$. The accumulated crossings (when normalized by $\alpha$) converge rapidly to the mean flow helicity in each case.

\begin{figure}[t!]
    \centering
  \includegraphics[width=.99\linewidth]{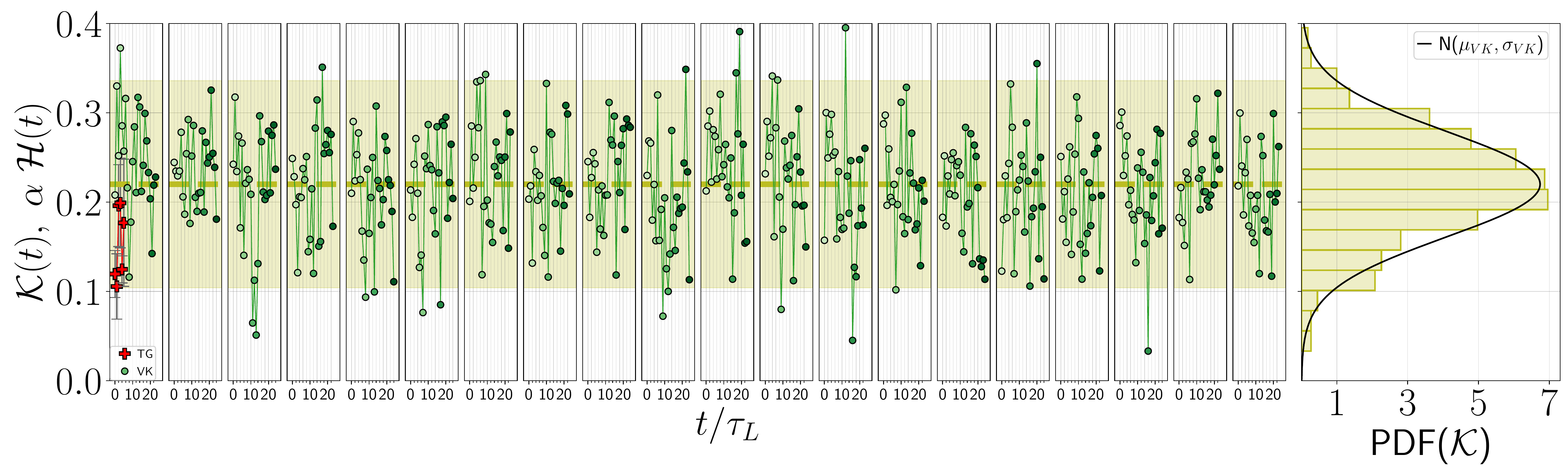}
    \caption{Time series of $\altmathcal{K}(t)$ for 20 realizations of the von K\'arm\'an experiment. Green dots turn darker as time evolves in each realization. In the first panel on the left, the time evolution of $\altmathcal{K}(t)$ for one TG cell with positive helicity is shown as a reference. On the right we show the PDF of $\altmathcal{K}$, with a normal distribution with same mean and dispersion as a reference.}
    \label{fig:K_VK}
\end{figure}

\subsection{Time dependence and uncertainties of $\altmathcal{K}$ and $\altmathcal{H}$}

In the DNSs, the large number of available trajectories allows for the determination of uncertainties in $\altmathcal{K}$ using different subsets of trajectories for the same time interval $\Delta T$: as was shown in the main text, for a fixed $\Delta T$ we can compute $\altmathcal{K}$ for subsets of the trajectories, and the error in $\altmathcal{K}$ is associated to the standard deviation of the subsets' values of $\LN$. However, in experiments, where the typical number of trajectories ranges from several tens to a few hundreds, to quantify uncertainties in $\altmathcal{K}$ we repeated the experiment several times. For each realization we then computed $\altmathcal{K}(t)$ and its average. Figure \ref{fig:K_VK} shows $\altmathcal{K}(t)$ as a function of time (in units of $\tau_L$) for 20 realizations of the von K\'arm\'an experiment, computed over non-overlapping time intervals $\Delta T = \tau_L$. The dashed line and the shaded region represent respectively the mean value over all realizations, and 95\% confidence levels. Within each realization $\altmathcal{K}(t)$ displays fluctuations, whose amplitudes are comparable to those in the TG DNSs (see the points corresponding to $\altmathcal{K}(t)$ for one TG cell with positive helicity in the first panel from the left in Fig.~\ref{fig:K_VK}). From all the realizations we can build the PDF shown in the rightmost panel of Fig.~\ref{fig:K_VK}. The uncertainty in the determination of $\altmathcal{K}$ for the experiment steady state is associated with the dispersion of this distribution.

In the LEM experiment, $\altmathcal{H}$ must be zero on average as the forcing mechanism is mirror symmetric. In the VK experiment, detailed studies (see, e.g., \cite{Angriman_2020}) have shown that the large-scale geometry of the VK flow resembles that of the TG flow in one cell, except for boundary effects. Thus, we estimate $\altmathcal{H}$ in this flow from calibrated DNSs of TG flows at similar Reynolds numbers, and we rescale values from the DNSs taking into account that the TG flow displays a different ratio of poloidal to toroidal velocities, $v_z/v_x$. From geometric considerations, $\altmathcal{H}$ in the experiment is thus estimated from $\altmathcal{H}$ in the DNSs rescaled by a factor $(v_z/v_x)_\text{VK} \, (v_z/v_x)^{-1}_\text{TG} \approx 1.15$.

\end{document}